\documentstyle[epsfig,aps,floats]{revtex}
\begin{document}
\tighten
%
%
\title{Formation of superheavy elements in heavy-ion collisions}

%
\author{V.Yu. Denisov$^{1,2}$}
\address{
$^1$ Gesellschaft f\"ur Schwerionenforschung, D-64291
Darmstadt, Germany \\
$^2$ Institute for Nuclear Research, 03680 Kiev, Ukraine}

%
%
\maketitle

\begin{abstract}

The cold fusion reactions related to $^{208}$Pb and $^{209}$Bi 
targets leading to superheavy elements (SHE) with Z=104-112 have been 
successfully considered in our model recently \cite{dh}. Here we 
briefly discuss this model and extend our consideration to fusion 
reactions between similar target and projectile. The reactions 
between weakly deformed target close to Pb and $^{76}$Ge, $^{82}$Se 
projectiles are also studied. The available experimental 
cross-sections are well described. The nucleus-nucleus interaction 
potential for reactions leading to SHEs are shortly discussed.
\end{abstract}

%
%
\section{Introduction}
The synthesis of superheavy elements (SHEs) was and still is an
outstanding research object. The properties of SHEs are studied
both theoretically and experimentally [1-7]. In the cold fusion, SHEs 
are produced by reactions of the type X+(Pb,Bi) $\rightarrow$ SHE+1n at 
subbarrier energies [1]. The excitation energy of a compound nucleus 
formed by cold fusion is low, $\approx$ 10--20 MeV [1]. The 
experimental study of an excitation function for the SHE production 
becomes increasingly difficult due to very small cross-sections and 
narrow width of the excitation function [1].

\begin{figure}
~\hspace{.5cm}\epsfig{file=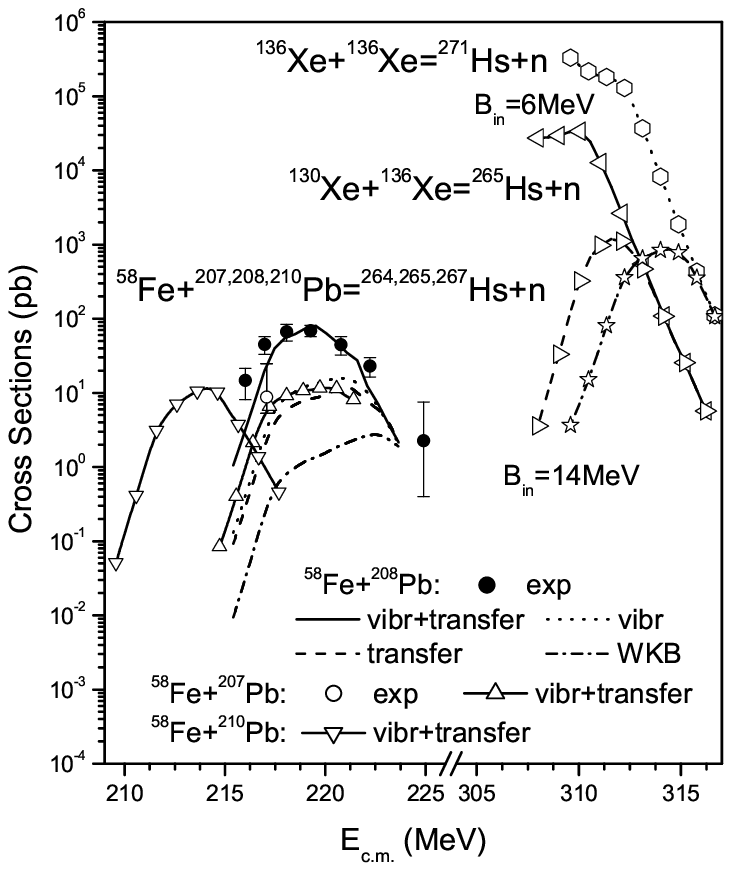,width=7.5cm}~\hspace{-.5cm}\epsfig{file=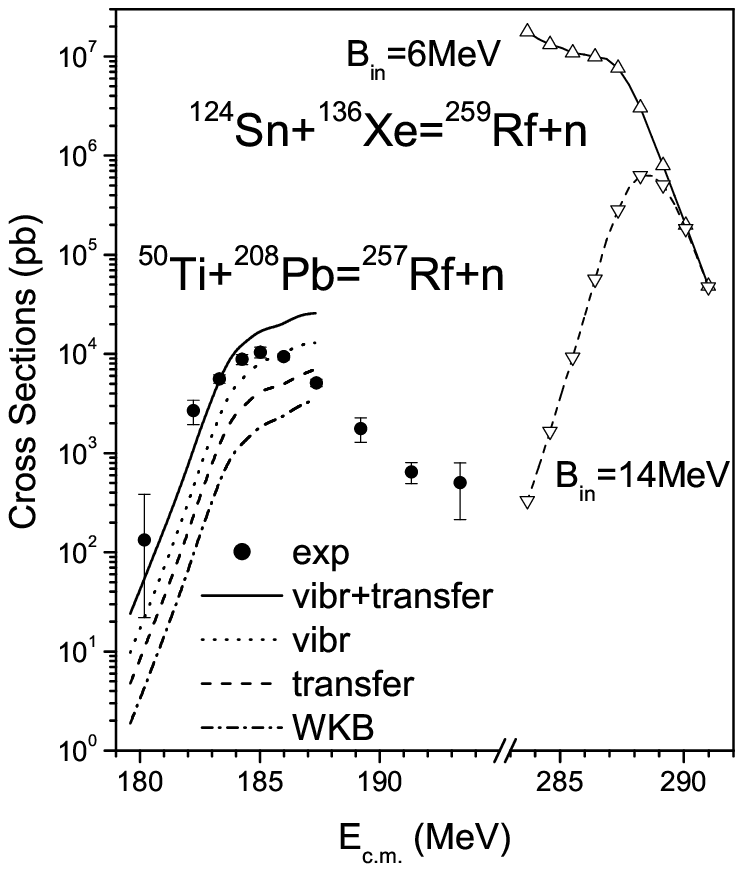,width=7.5cm}
\caption{Calculated excitation functions for reactions  
$^{58}$Fe + $^{207,208,210}$Pb $\rightarrow ^{264,265,267}$Hs + n (left),
$^{130,136}$Xe + $^{136}$Xe$\rightarrow ^{265,271}$Hs + n (left),
$^{50}$Ti + $^{208}$Pb$\rightarrow ^{257}$Rf + n (right)
and $^{124}$Sn+$^{136}$Xe$\rightarrow ^{259}$Rf + n (right).  
For reactions 
$^{58}$Fe + $^{208}$Pb$\rightarrow ^{265}$Hs + n and
$^{50}$Ti + $^{208}$Pb$\rightarrow ^{257}$Rf + n the solid 
curves take into account 
both the low-energy $2^+ $ and $3^-$ vibrations and the 
neutrons transfer channels. The dotted and the dashed curves for
these reactions show the results based on solely the $2^+ $ and $3^-$ 
vibrations and the neutron transfer channels, respectively. The results 
of the one-dimensional WKB approach for these reactions are 
shown by the dash-dotted curves. Calculations including 
both vibrations and transfer enhancements obtained for $^{130,136}$Xe+$^{136}$Xe$\rightarrow ^{265,271}$Hs+n and
$^{124}$Sn+$^{136}$Xe$\rightarrow ^{259}$Rf+n reactions 
are additionally marked by symbols, see assignments. The 
experimental data are taken from [1,4].}
\end{figure}

\begin{figure}
~\hspace{1cm}\epsfig{file=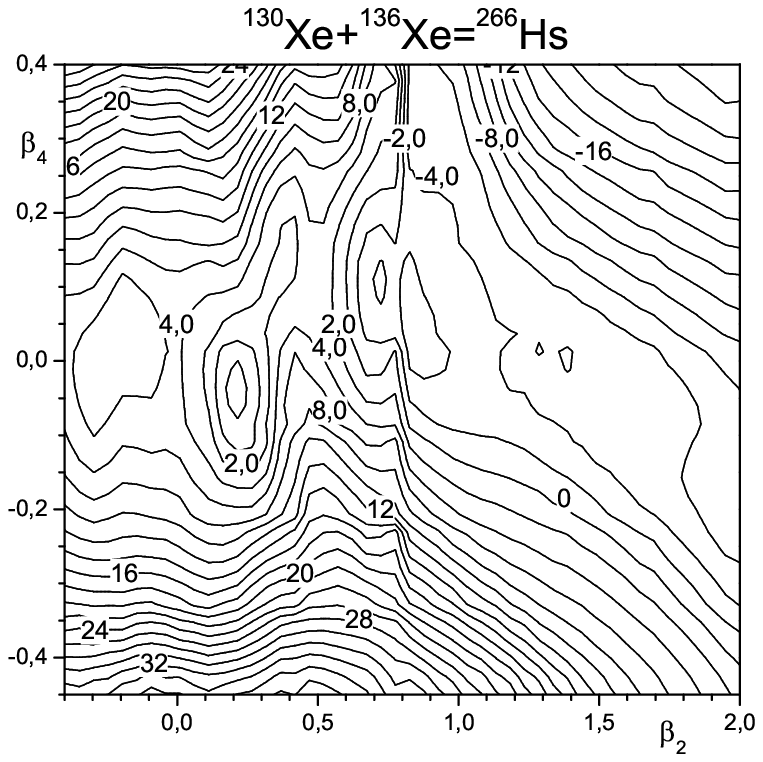,width=8.5cm}~\hspace{-.5cm}\epsfig{file=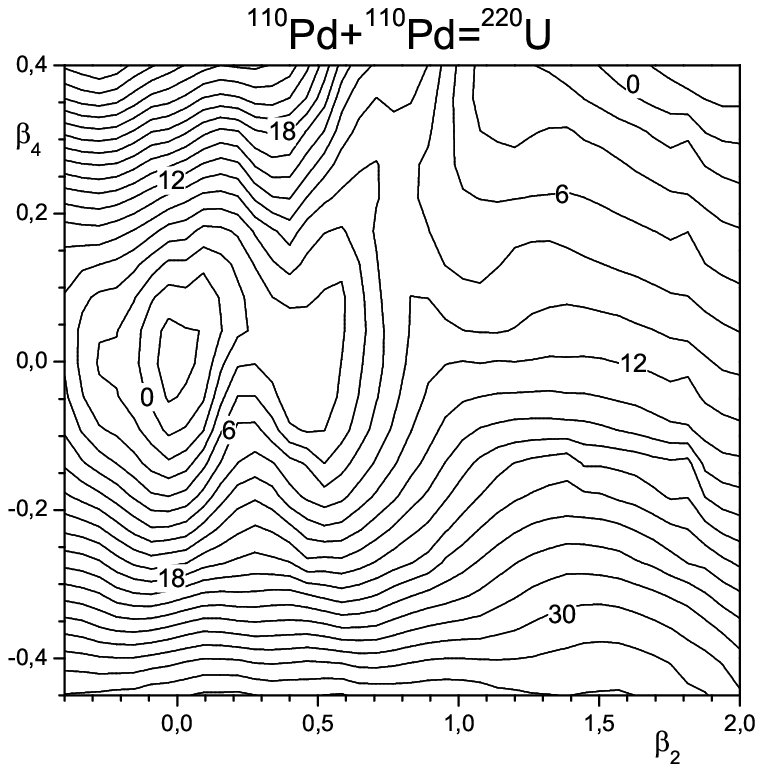,width=8.5cm}
\caption{Potential energy surfaces as a function of the 
deformation parameters $\beta_2$ and $\beta_4$ for cold 
fusion reactions $^{130}$Xe+$^{136}$Xe$\rightarrow ^{266}$Hs (left)
and $^{110}$Pd+$^{110}$Pd$\rightarrow ^{220}$U (right). 
Parameters $\beta_\ell$ are equal to:  $\beta_2=p \beta_2^t$, 
$\beta_4=q \beta_4^t$. The touching configurations of the spherical
projectile and target nuclei are close to the bottom right corner or 
each map and that of the ground-states are close to $\beta_2 \approx 
0.2 (0.0)$ and $\beta_4 \approx -0.05 (0.0)$ for $^{266}$Hs 
($^{220}$U). The contour lines are drawn every 2 MeV.\label{fig2}}
\end{figure}

In Refs. \cite{dh,d} we present a model for description of
measured excitation functions for the SHE production in the
cold fusion reactions. The maximum position and the
width of the excitation function for cold fusion reactions
X+$^{208}$Pb,$^{209}$Bi leading to elements with Z=104-112 are
well described in \cite{dh} (see also Figs. 1,3,4). 
Within our approach \cite{dh} the process of the SHE formation proceeds 
in three stages: ({\sl i}.) 
The capture of two spherical nuclei and the formation of a
common shape of the two touching nuclei. Low-energy surface
vibrations and a transfer of few nucleons are taken into account
at the first step of the reaction. ({\sl ii}.)
The formation of a spherical or near spherical compound nucleus.
({\sl iii}.) The surviving of an excited compound
nucleus during evaporation of neutrons and $\gamma$-ray emission
which compete with fission. A reduction of the fission barrier
is taken into account, which arises from a reduction of the shell
effects at increasing excitation energy of the compound nucleus.

One of the heaviest systems experimentally studied over a wider
range of excitation energy is $^{58}$Fe+$^{208}$Pb $\rightarrow
^{265}$Hs+n \cite{hm}, the data are shown in Fig. 1. The experimental
data are compared with several modifications of our model.
In the simplest case, using tunneling through a one-dimensional 
barrier and the WKB method, the model strongly underestimates the
experimental fusion cross-sections. Better agreement is obtained,
when the neutron transfer channels from lead to iron are taken
into account. Similarly, the cross-sections increase if the 
low-energy $2^+$ and $3^-$ surface vibrational excitations 
of both projectile and target are included in the calculations, 
see Fig. 1. The best fits are obtained by considering transfer 
and vibrations simultaneously. The values of parameters and other
details are presented in Ref. \cite{dh}. In our model \cite{dh} we have 
two fitting parameters as well as other parameters, which are taken 
from experimental data and from other calculations, see for detail 
\cite{dh}. Note that we are able to describe data for reactions 
$^{58}$Fe+$^{207}$Pb $\rightarrow ^{264}$Hs+n (see Fig. 1) and 
$^{58}$Fe+$^{209}$Bi $\rightarrow ^{266}$Mt+n 
(see Fig. 11 in \cite{dh}) by using the same fitting parameters as
fixed for reaction $^{58}$Fe+$^{208}$Pb $\rightarrow ^{265}$Hs+n.

The results of similar calculations for reactions $^{50}$Ti+$^{208}$Pb 
$\rightarrow ^{257}$Rf+n, $^{62,64}$Ni+$^{207,208,210}$Pb $\rightarrow $110+n, and $^{54}$Cr+$^{208}$Pb $\rightarrow ^{261}$Sg+n are also presented in Figs. 1,3,4. Our calculation for reaction $^{50}$Ti+$^{208}$Pb $\rightarrow ^{257}$Rf+n 
is well agreed with the experimental data for low collision energies, see Fig. 1. Note that the threshold of $2n$ evaporation channel is near 
186 MeV. We have not take into account the emission of the second 
neutron, which should reduce the difference between our calculations 
and experimental data for energies close to 186-188 MeV.

\section{Production of SHE in Nearly Symmetric Heavy Ion Collisions}

It is also possible to produce SHE in collisions of similar nuclei.
For example, the nuclide $^{265}$Hs can be formed in both reactions
$^{58}$Fe+$^{208}$Pb $\rightarrow ^{265}$Hs+n and 
$^{130}$Xe+$^{136}$Xe $\rightarrow ^{265}$Hs+n. 
The excitation functions for reactions  
$^{130,136}$Xe+$^{136}$Xe $\rightarrow ^{265,271}$Hs+n 
obtained in our model for different values of the inner barrier 
$B_{\rm Sph} = 6 $ MeV and 14 MeV are presented in Fig. 1. The inner 
barrier is the barrier appearing during the shape evolution from two 
touching nuclei to the compound nucleus [2,3]. The shape evolution stage is related with the second step of the SHE formation.

We can describe the touching configuration and shape evolution
of nearly symmetric systems by using the shape parametrization 
$ R(\vartheta) = R(p,q) [1 + p \sum_{\ell=2, \ell \ne 4}^9 \beta_\ell^t 
Y_{\ell 0} (\vartheta)+q \beta_4^t Y_{4 0} (\vartheta) ] $ , 
where $p=q=1$ at the touching point of two spherical colliding ions 
and the deformation parameters $\beta_\ell^t$ are fixed at the
touching point. This parametrization of the nuclear shape near 
touching point is very accurate for symmetric case. The volume of the 
nucleus is constant during the shape evolution. The potential energy 
surfaces for reactions $^{130}$Xe+$^{136}$Xe$\rightarrow 
^{266}$Hs and $^{110}$Pd+$^{110}$Pd$\rightarrow 
^{220}$U are presented in Fig. 2. 

As mentioned above, we made calculations of synthesis
of the SHE using nearly symmetric reactions for two different 
values of the inner barrier. The value of the inner barrier 
$B_{\rm Sph} = 6 $ MeV is near the adiabatic fission
barrier, see Fig. 2. Note that the potential energy surface for 
$^{266}$Hs in Fig. 2 has a strong slope from touching point 
(bottom right corner) to the quasi-fission direction (upper right
corner). It is possible to evaluate exact value of the 
inner barrier and the branching ratio between compound nucleus 
formation and quasi-fission processes by studying the quantum dynamical 
shape evolution from two touching nuclei to the near spherical compound 
nucleus. Unfortunately such calculations are not available now. 
Therefore, we also made calculations for a higher value of the 
inner barrier $B_{\rm Sph} = 14 $ MeV, which may simulate 
a stronger competition between the SHE formation and the quasi-fission. 
Note that the SHE production cross sections for near symmetric
reactions in the case of $k \geq 1$ neutrons evaporation are mainly 
related to the shape evolution stage, because the height of the outer 
(capture) barrier is smaller then collision energies.

Nucleus $^{220}$U is the heaviest nucleus, which have been formed in 
symmetric reaction $^{110}$Pd + $^{110}$Pd $\rightarrow ^{220}$U  
\cite{pdpd}. The landscape of potential energy surface for Pd+Pd  
reaction has much stronger repulsion at touching point region
then the one for Xe+Xe reaction, see fig. 2. Due to this the fusion 
trajectory for Pd+Pd reaction at low energies should strongly deviates 
to $\beta_4 \approx 0$ near touching point or penetrates through the 
thick barrier near $\beta_2 \approx 1.0 \div 1.7$ and $\beta_4 \approx 
-0.4 \div -0.25$. So the SHE production is reduced in both cases. 
Nevertheless  the cross section for reaction $^{110}$Pd + $^{110}$Pd 
$\rightarrow ^{220}$U at lowest energy close to 1 nb \cite{pdpd}. We 
may expect that near symmetric reactions as Sn+Xe or Xe+Xe very 
attractive tools to produce SHEs, because they have thin barrier near 
$\beta_2 \approx 0.5 \div 0.8$, $\beta_4 \approx -0.4 \div 0.05$ and absence of strong repulsion near the touching point. 

\section{Reaction with Ge and Se projectiles}

\begin{figure}
\begin{center}
\epsfig{file=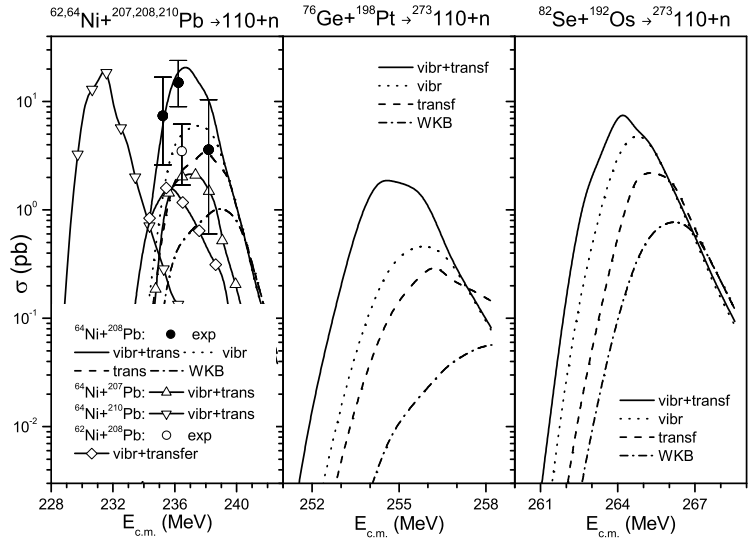,width=10.9cm}
\caption{Calculated excitation functions for the reactions 
$^{62,64}$Ni+$^{207,208,210}$Pb $\rightarrow $110+n (left),
$^{76}$Ge+$^{198}$Pt $\rightarrow ^{273}$110+n (middle)
and $^{82}$Se+$^{192}$Os $\rightarrow ^{273}$110+n (right).  
The solid curve takes into account 
both the low-energy $2^+ $ and $3^-$ vibrations and the 
neutron transfer channels. The dashed, dotted and dash-dotted
lines show the results of similar calculations as for reaction $^{58}$Fe+$^{208}$Pb $\rightarrow ^{265}$Hs+n in Fig. 1. The experimental data are taken from [1].}
\end{center}
\end{figure}

Reactions with targets slightly lighter then Pb are also interesting
for studying too. The reactions between
$^{76}$Ge+$^{198}$Pt $\rightarrow ^{273}$110+n, 
$^{82}$Se+$^{192}$Os $\rightarrow$ $^{273}$110+n ,
$^{76}$Ge+$^{186}$W $\rightarrow ^{261}$Sg+n and 
$^{82}$Se+$^{180}$Hf $\rightarrow ^{261}$Sg+n are compared with the 
cold fusion reactions leading to the same SHE in Figs. 3,4. 

Note that our code is proposed for the evaluation of the SHE production 
cross section in fusion of two spherical nuclei \cite{dh}. However 
we may simulate static deformation effect in deformed target by taking 
the same value of the vibration $2^+$ amplitude as the value of static 
quadrupole deformation. We also equate the energy of the first $2^+$ 
vibration state to the value of the first rotational $2^+$ state in 
deformed nuclei.

Reactions with $^{76}$Ge and $^{82}$Se projectiles in Figs. 3,4 are 
strongly enhanced by coupling to both neutron transfer and 
low-energy surface vibrations. More symmetric reactions with $^{82}$Se 
projectile have smaller difference between the capture barrier height 
and the ground state $q$-value energy of compound nucleus formed at 
fusion. Due to this, the SHE production cross sections for reactions 
with Se projectile are higher then the one for Ge case, see Figs. 3,4. 

\begin{figure}
\begin{center}
\epsfig{file=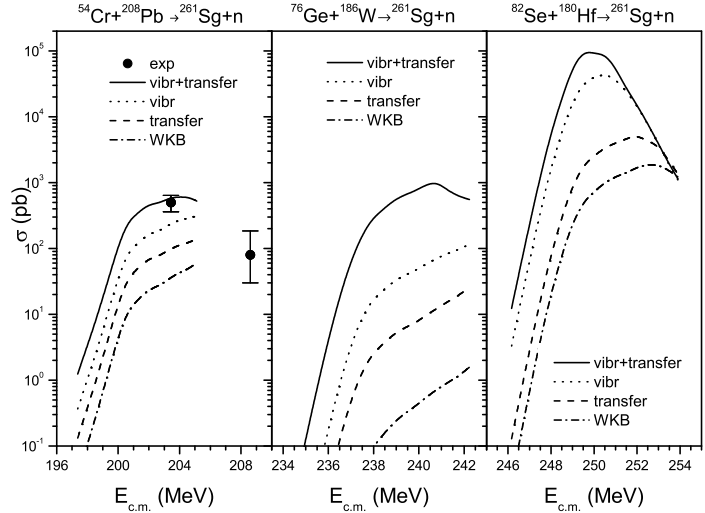,width=10.2cm}
\end{center}
\caption{Calculated excitation functions for the reactions 
$^{54}$Cr+$^{208}$Pb $\rightarrow ^{261}$Sg+n (left),
$^{76}$Ge+$^{186}$W $\rightarrow ^{261}$Sg+n (middle)
and $^{82}$Se+$^{180}$Hf $\rightarrow ^{273}$Sg+n (right).  
The line assignments are the same as in Fig. 3. 
The experimental data are taken from [1].}
\end{figure}

\section{Interaction potential between very heavy nuclei}

The ion-ion potential near and before the touching point is quite 
differently treated in models [2-3,5-7]. Note that the interaction 
potential between very heavy ions near the touching point is not well known, because there are not so many experimental data for such cases. 
The knowledge of the ion-ion potential near touching point is crucial 
for the capture stage of the SHE production. The uncertainty of the 
interaction potential between heavy ions near the touching point gives 
possibility for very different proposals on the mechanisms of the SHEs 
formation [2-3,5-7]. So, there is necessity to reduce the uncertainty 
of the interaction potential values near the touching point.

\begin{figure}
\begin{center}
\epsfig{file=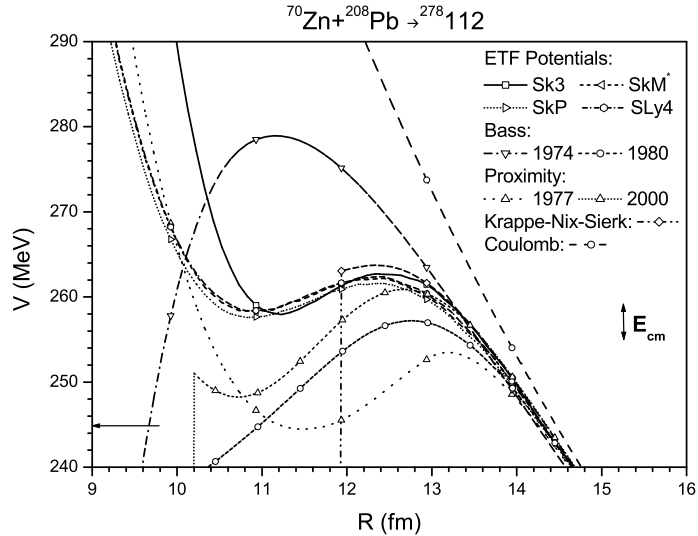,width=10.2cm} 
\end{center}
\caption{The potentials for the collision $^{70}$Zr+$^{208}$Pb
evaluated in the ETF approximation with Sk3, SkM$^*$, SkP and SLy4
parameter sets of the Skyrme force are given. The Coulomb potential and
the potentials obtained in the two versions (1974 and 1980 years) of
the Bass parametrization [11], in the KNS parametrization [12] and in 
two versions of proximity approximation (1977 and 2000 years) [9,10] 
are also presented. The range of collision energies in the middle of 
the target used in experiments [1] is marked by vertical arrow in the 
right part. The ground state reaction Q-value related with the 
formation of compound nucleus $^{278}112$ is marked by the horizontal 
arrow in the left bottom corner.}
\label{fig:1}       
\end{figure}

We calculate the interaction potential between heavy ions around the 
touching point for various Skyrme forces in the extended Thomas-Fermi 
(ETF) approximation by using the frozen Hartree-Fock-Bogoliubov 
densities of the individual nuclei \cite{dn}. In Fig. 5 we present the 
interaction potentials between $^{70}$Zr and $^{208}$Pb evaluated in 
the ETF approximation with different parameter sets of the Skyrme 
force. The potentials obtained by using different analytical 
expressions \cite{prox77,prox2000,bass,kns} are also shown in Fig. 5.
The potentials obtained in the ETF frozen approximation with different 
parameter sets of the Skyrme force have very close values of both the 
barrier heights and the barrier distances. The ion-ion potentials 
evaluated for SkM$^*$, SkP and SLy4 sets are very close to each other 
at all distances presented in Fig. 5. However Sk3 parameter set of the 
Skyrme force has largest value of the compressibility among all 
considered here sets of the Skyrme force. Due to this the repulsion 
between ions at small distances related with the volume of overlapped 
ions densities is strongest in the case of Sk3 set of the Skyrme force.

The potential wells obtained in the ETF approximation are shallow, see 
Fig. 5. The minimal value of the potential well is located at the 
distance smaller then the touching point distance of two spherical 
ions, which is close to 12 fm. Here we roughly determined touching 
point distance as $R=r_0(A_1^{1/3}+A_2^{1/3})$, where $r_0=1.2$ fm. 
Therefore both processes, the capture of two ions and the neck 
formation, are taken place in the potential well. The barriers obtained 
with the help of different analytical expressions for ion-ion potential 
\cite{prox77,prox2000,bass,kns} are spread in very wide interval in 
Fig. 5. 

The further studies shows that the potentials between light and medium 
heavy ions have deep and wide capture well. The depth of these well 
decrease with increasing $Z_1 Z_2$ and disappear for very heavy 
colliding ions \cite{dn}. Due to this the capture process is suppressed 
for the case of reactions between extremely heavy ions.

Concluding, we note that it would be interesting to check 
experimentally our estimates for both symmetric reactions and 
reactions with $^{76}$Ge and $^{82}$Se projectiles. 

The author would like to thank S. Hofmann, V. Ninov, W. N\"orenberg and
J. Peter for useful discussions. He acknowledges gratefully support 
from ISOL'2001 Organizing Committee of the Conference and GSI.

\bibliographystyle{try}

%
%
\end{document}